\documentstyle[emulateapj5,12pt,onecolfloat]{aastex}
\def\arcs{$''$}

\begin{document}
\twocolumn[
\title{Star Formation at $z\sim6$: The UDF-Parallel ACS
Fields\altaffilmark{1}}

\author{R.J. Bouwens$^{2}$, G.D. Illingworth$^{2}$,R.I. Thompson$^{3}$,
        J.P. Blakeslee$^{4}$,M.E. Dickinson$^{5}$,T.J. Broadhurst$^{6}$,
        D.J. Eisenstein$^{3}$, X. Fan$^{3}$,M. Franx$^{7}$,G. Meurer$^{4}$,
        P. van Dokkum$^{8}$}
\affil{1 Based on observations made with the NASA/ESA Hubble Space
Telescope, which is operated by the Association of Universities for
Research in Astronomy, Inc., under NASA contract NAS 5-26555. These
observations are associated with programs \#9803.}
\affil{2 Astronomy Department, University of California, Santa Cruz,
CA 95064}
\affil{3 Steward Observatory, University of Arizona, Tucson, AZ 85721.}
\affil{4 Department of Physics and Astronomy, Johns Hopkins
University, Baltimore, MD 21218.}
\affil{5 National Optical Astronomy Obs., P.O. Box 26732, Tucson, AZ 85726.}
\affil{6 Racah Institute of Physics, The Hebrew University, Jerusalem,
Israel 91904.}
\affil{7 Leiden Observatory, Postbus 9513, 2300 RA
Leiden, Netherlands.}
\affil{8 Department of Astronomy, Yale University, New Haven, CT 06520.}

%\author{Selected members of the UDF NICMOS team}

%\and 

%\author{Selected members of the ACS GTO team}

\begin{abstract}
We report on the $i$-dropouts detected in two exceptionally deep ACS
fields ($B_{435}$, $V_{606}$, $i_{775}$, and $z_{850}$ with $10\sigma$
limits of 28.8, 29.0, 28.5, and 27.8, respectively) taken in parallel
with the UDF NICMOS observations.  Using an $i-z>1.4$ cut, we find 30
$i$-dropouts over 21 arcmin$^2$ down to $z_{850,AB}=28.1$, or 1.4
$i$-dropouts arcmin$^{-2}$, with significant field-to-field variation
(as expected from cosmic variance).  This extends $i$-dropout searches
some $\sim$0.9$^m$ further down the luminosity function than was
possible in the GOODS field, netting a $\sim$7$\times$ increase in
surface density.  An estimate of the size evolution for UV bright
objects is obtained by comparing the composite radial flux profile of
the bright $i$-dropouts ($z_{850,AB}<27.2$) with scaled versions of
the HDF-N + HDF-S $U$-dropouts.  The best-fit is found with a
$(1+z)^{-1.57_{-0.53} ^{+0.50}}$ scaling in size (for fixed
luminosity), extending lower redshift ($1<z<5$) trends to $z\sim6$.
Adopting this scaling and the brighter $i$-dropouts from both GOODS
fields, we make incompleteness estimates and construct a $z\sim6$ LF
in the rest-frame continuum UV ($\sim1350\AA$) over a 3.5 magnitude
baseline, finding a shape consistent with that found at lower
redshift.  To evaluate the evolution in the LF from $z\sim3.8$, we
make comparisons against different scalings of a lower redshift
$B$-dropout sample.  Though a strong degeneracy is found between
luminosity and density evolution, our best-fit model scales as
$(1+z)^{-2.8}$ in number and $(1+z)^{0.1}$ in luminosity, suggesting a
rest-frame continuum $UV$ luminosity density at $z\sim6$ which is just
$0.38_{-0.07} ^{+0.09}\times$ that at $z\sim3.8$.  Our inclusion of
size evolution makes the present estimate lower than previous $z\sim6$
estimates.
\end{abstract}
]
\section{Introduction}

The unique deep $z$-band capabilities of the Hubble Space Telescope
(HST) Advanced Camera for Surveys have uncovered a substantial
population of $z\sim6$ dropouts (e.g., Yan, Windhorst, \& Cohen 2003).
Early estimates indicate that the star formation rate at $z\sim6$ is
only somewhat lower than (only 0.17-0.77$\times$ that) at $z\sim3$
(Bouwens et al.\ 2003b; Stanway, Bunker, \& McMahon 2003; Dickinson et
al.\ 2004; Stanway et al.\ 2004; Bouwens et al.\ 2004, hereinafter,
B04).  Unfortunately, these early studies have all been on fields of
somewhat intermediate depth, with strong incompleteness corrections
beyond $z_{850,AB}=26.5$.  Since a simple extrapolation of the Steidel
et al.\ (1999) $z\sim3$ luminosity function to $z\sim6$ predicts an
$L_{*}$ of $z_{850,AB}=26$, these samples are becoming incomplete just
as the $i$-dropouts are starting to become prominent.  Moreover, the
substantial biases against larger, lower surface brightness objects
make it difficult to derive the size distribution at $z\sim6$ (e.g.,
B04 find that only $\sim$25\% of the brighter $U$-dropouts make it
into their $z\sim6$ Great Observatories Origins Deep Survey (GOODS)
sample).  It is therefore important to examine $i$-dropouts in a much
deeper field.

Here we report on observations of $i$-dropouts from two exceptionally
deep ACS Wide Field Camera (WFC) parallels (hereinafter, UDF-Ps) that
were taken in conjunction with Near Infrared Camera and Multi-Object
Spectrometer (NICMOS) observations of the Ultra Deep Field (UDF)
(GO-9803; Thompson et al.\ 2004).  9, 9, 18, and 27 orbits were taken
in the F435W, F606W, F775W, and F850LP passbands (hereafter called
$B_{435}$, $V_{606}$, $i_{775}$, and $z_{850}$) for each field over 9
pointings (the pointings were arranged in a $3\times3$ grid, each
separated by 45\arcs).  These observations extend $i$-dropout searches
some 0.9$^m$ magnitude deeper than GOODS and to within 0.9$^m$
magnitudes of the UDF, giving a preview of the results at these
depths.  We assume $(\Omega_M,\Omega_{\Lambda},h) = (0.3,0.7,0.7)$
(Bennett et al.\ 2003).

\section{Observations}

These observations were retrieved from STScI and subject to the
standard CALACS on-the-fly reprocessing (OTFR).  Additional
processing, including cosmic-ray rejection, alignment, and final image
combination, was performed by the ``Apsis'' data reduction software
(Blakeslee et al. 2003).  The latest photometric zero points were
applied (Sirianni et al.\ 2004) along with a correction for galactic
absorption (Schlegel, Finkbeiner, \& Davis 1998).  The $10\sigma$
limiting magnitudes (for the deepest parts of the image) were
approximately 28.8, 29.0, 28.5, and 27.8 in the $B_{435}$, $V_{606}$,
$i_{775}$, and $z_{850}$ bands, respectively, with PSF FWHMs of
0.08-0.09\arcs.

\section{Analysis}

Areas whose $V$, $i$, and $z$ exposures were shorter than 5, 10, and
15 orbits were excluded from the analysis.  Object detection and
photometry was performed on the remaining area for each parallel (21
arcmin$^2$, in total) using SExtractor (Bertin \& Arnouts 1996) with
smaller scaled apertures (MAG\_AUTO) to measure colors and larger
scaled apertures to correct to total flux.  For our $i$-dropout
selection, we apply a similar $i$-dropout selection criteria to that
used by B04 on the GOODS data set.  We require objects be a 5.5
$\sigma$-detection within a 0.25$''$-diameter aperture, have a
SExtractor stellarity less than 0.85 (non-stellar to high confidence),
have $(i_{775}-z_{850})_{AB}$ colors redder than 1.4, and be a
non-detection ($<2\sigma$) in the $V_{606}$ band (within the
smaller-scaled apertures).  Several studies have demonstrated that a
simple $i-z$ cut can be an effective means of selecting $z>5.5$
galaxies (Stanway et al.\ 2003; Bouwens et al.\ 2003b; B04; Dickinson
et al.\ 2004).  Since the surface densities of all known interlopers
are expected to increase less quickly than bona-fide $i$-dropouts, we
adopt the 11\% found for brighter $i$-dropouts from GOODS (B04) as an
upper limit to the contamination rate, with no spurious detections
(running through a similar selection procedure on the negative images,
no $i$-dropouts were found.)  Given the similar incompleteness
expected due to blending with foreground objects (verified through
simulations, see B04), no net adjustment is made to the surface
density.

\textit{(a) Surface Density.} With the above selection criteria, we
recovered 30 $i$-dropouts to a limiting $z_{850,AB}$ magnitude of
28.1, 20 in the first parallel and 10 in the second, suggesting a
significant degree of clustering, but not inconsistent with the cosmic
variance (50\% RMS) expected for fields of this size (Somerville et
al.\ 2004; B04).  The luminosities of $i$-dropouts in our sample range
from $1.1\, L_{*}$ ($z_{850,AB}=25.9$) to $0.15\, L_{*}$
($z_{850,AB}=28.1$) (using the Steidel et al.\ (1999) value for
$L_{*}$ at $z\sim3$).  Half-light radii (uncorrected for PSF effects)
range from 0.1$''$ to 0.3$''$, or 0.6 kpc to 1.7 kpc for the adopted
set of cosmological parameters.  The positions, magnitudes, sizes,
$i-z$ colors, and SExtractor stellarity parameters of these objects
are tabulated in Table 1.

\textit{(b) Size Evolution.}  Cosmic surface brightness dimming
results in significant incompleteness at $z\sim6$ (Bouwens et al.\
2003b; Giavalisco et al.\ 2004; B04).  To treat this, it is necessary
to have a good estimate of the size (surface brightness) distribution
(independent of selection effects).  An empirical way of estimating
this distribution is to model the observed sizes in terms of a scaled
version of a lower redshift population and therefore derive the size
evolution.  For our low redshift comparison sample, we adopt the
$z\sim2.5$ HDF-N + HDF-S $U$-dropout sample (B04).  The large $\Delta
\log (1+z)$ increases our sensitivity to size changes.  To artifically
redshift this sample to $z\sim6$--weighting by $1/V_{max}$ and
accounting for pixel $k$-corrections, surface brightness dimming, and
detailed image properties like the noise and the PSF (as well as our
selection procedure given above)--we employ the now well-established
cloning machinery (Bouwens, Broadhurst, \& Silk 1998a,b; Bouwens et
al.\ 2003a,b; B04).  Finally, this redshifted sample is compared with
a sample of $i$-dropouts ($z_{850,AB}<27.2$) extracted from the ACS
images degraded to the PSF of our $U$-dropout images projected to
$z\sim6$ (0.14\arcs$\,$ FWHM) with a S/N equivalent to 5 orbit
$V_{606}$, 10 orbit $i_{775}$, and 15-orbit $z_{850}$ observations
($0.6^m$ deeper than GOODS: see sample \#2 from Table 1 for the
objects used here).

\begin{figure}[h]
\epsscale{0.95}
\plotone{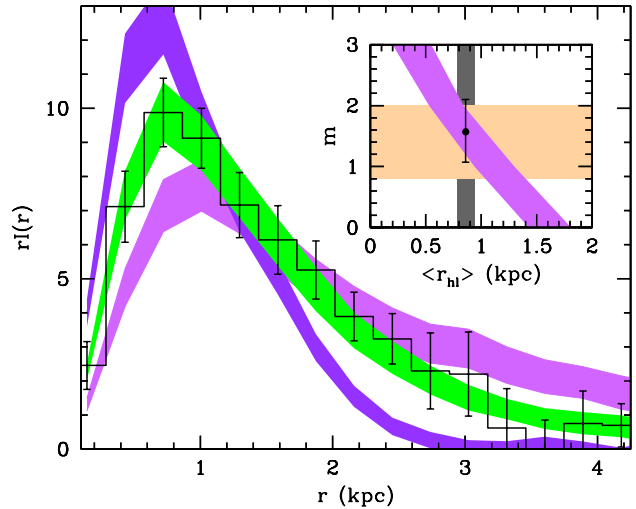}
\caption{The mean radial flux profile for the 10 brightest
$z_{850,AB}<27.2$ $i$-dropouts (histogram, with $\pm1\sigma$
uncertainties) in our sample versus the predictions based upon
different scalings of the $z\sim2.5$ HDF-N + HDF-S $U$-dropout sample
(the $i_{775}$ images available for this sample).  The shaded violet,
green, and dark violet regions represent predictions based upon the
$(1+z)^0$, $(1+z)^{-1.5}$, and $(1+z)^{-3}$ size scalings,
respectively.  Clearly, the $(1+z)^{-1.5}$ scaling is a much better
fit to the observations than the other two scalings.  This is
quantified in the figure inset, which shows the expected $1\sigma$
scatter (\textit{diagonal violet band}) in the mean half-light radius
vs. the $(1+z)^{-m}$ size scaling exponent $m$.  Since the half-light
radius for the mean profile is $0.85\pm0.06$ kpc (\textit{plotted as a
gray vertical band}), the inferred scaling is $(1+z)^{-1.57_{-0.50}
^{+0.53}}$ (shown as the solid black circle with $1\sigma$ error bars)
or $2.9_{-0.8} ^{+1.3} \times$ from $z\sim2.5$ to $z\sim5.9$ (the mean
redshifts of the $U$ and $i$-dropout samples, B04).  This is in good
agreement with the allowed range of scalings ($m=0.8-2.0$) inferred
from a recent analysis of the GOODS data (B04) (\textit{shaded in
light orange}).}
\end{figure}

We consider size scalings of the form $(1+z)^{-m}$ ($m=0-3$) (for
fixed luminosity) and quantify the differences in terms of the mean
radial flux profile, which provides a measure of the light within
different circular annuli (see B04 for the computational details).
The results of the procedure are shown in Figure 1 for the $(1+z)^0$,
$(1+z)^{-1.5}$, and $(1+z)^{-3}$ scalings, and it is obvious that
while the $(1+z)^{-1.5}$ scaling provides a good fit, the $(1+z)^0$
and $(1+z)^{-3}$ scalings produce profiles which are too broad and
narrow, respectively (see Figure 2 for a visual comparison of the
brighter dropouts with the no-evolution $m=0$ scalings).  To use these
results to place confidence limits on the allowed scalings, we
computed the mean size and variance for a 10-object sample of
intermediate magnitude $i$-dropouts.  The result is shown in the inset
to Figure 1 as the shaded violet region.  Given the observed
half-light radius $0.85\pm0.06$ kpc (corrected for PSF effects)
(\textit{shaded grey region}), we infer an $m$ of $1.57_{-0.50}
^{+0.53}$ (\textit{shown in the inset as the solid circle with error
bars}), similar to what was found from our GOODS study (\textit{light
orange region}) where $m$ was found to be between 0.8 and 2.0 (B04).

\textit{(c) Luminosity/Density Evolution.}  To interpret the
$i$-dropout counts and therefore estimate the evolution of the LF, we
use different projections of the lower redshift $B$-dropout sample
from the GOODS fields (B04).  We explore scalings in both luminosity
(e.g., $(1+z)^l$) and number (e.g., $(1+z)^n$), assessing these
scalings by comparing against both the UDF-Ps (degraded to the
$B$-dropout PSF projected to $z\sim6$, 0.1\arcs$\,$ FWHM with a S/N
equivalent to 5 orbit $V_{606}$, 10 orbit $i_{775}$, and 15 orbit
$z_{850}$ observations: see sample \#3 from Table 1 for the objects
used here) and the GOODS fields (B04) (corrected up to the same
completeness level as that of the UDF-Ps), which provide important
constraints at bright magnitudes.  As per the results of the previous
subsection, sizes of individual $B$-dropouts are scaled as
$(1+z)^{-1.5}$ (increasing their surface brightness).  Interpolating
between the simulation results and calculating a reduced $\chi^2$ for
each comparison, the 68\% and 95\% likelihood contours can be
determined, taking as inputs the GOODS $i$-dropouts alone
(\textit{shown in the Figure 3 inset as the green contours}) and GOODS
combined with the UDF-Ps (\textit{black contours}).  While the results
from the UDF-Ps are helpful in breaking the degeneracy between
luminosity and density evolution, small differences in normalization
between the bright (from GOODS) and faint (from the UDF-Ps) counts
could affect the results (due to cosmic variance).  Deep wide-field
images will be needed to ultimately establish the form of the
evolution.  The $(1+z)^{-2.8}$ scaling in number and $(1+z)^{0.1}$
scaling in luminosity preferred in this analysis point towards a
$z\sim6$ luminosity density ($\sim1350\AA$) that is just $0.38_{-0.07}
^{+0.09}\times$ that at $z\sim3.8$ (error bars computed from 2D
likelihood contours).  Applying this factor to the $z\sim3.8$
rest-frame continuum $UV$ luminosity density determined by B04 and
using standard assumptions to convert this into a star formation rate
density (e.g., Madau, Pozzetti, \& Dickinson 1998), the star formation
history integrated down to $0.2$ $L_{*}$ can be plotted and compared
with several previous determinations (Figure 4).  The present result
is noticeably lower.  This is almost entirely the effect of size
(surface brightness) evolution (see Figure 1) on the estimated
completeness of the $z\sim6$ census (and not due to our greater
depth).  Without size evolution from $z\sim3.8$, the completeness
correction and inferred luminosity density are some $\sim1.9\times$
higher than the current result (closer to previously published
estimates: Bouwens et al.\ 2003b; Giavalisco et al.\ 2004; B04--see
Figure 4).

\begin{figure}[t]
\twocolumn[
\epsscale{1.8}
\plotone{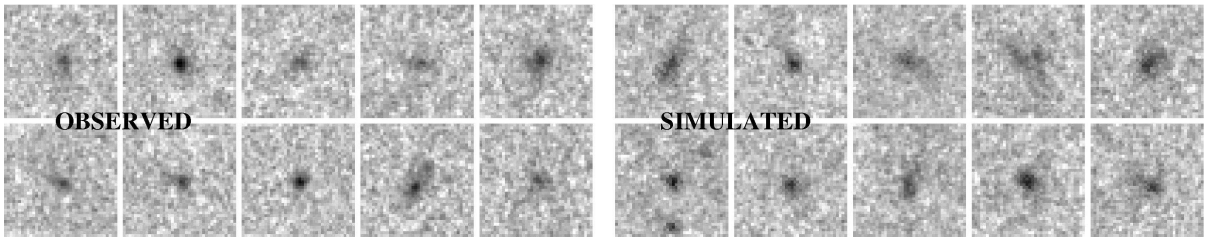}
\caption{Postage stamp images ($z_{850}$-band) of 10 randomly-selected
$i$-dropouts from the UDF-Ps ($z_{850,AB}<27.2$) versus similarly
selected no-evolution projections of our lower redshift HDF
$U$-dropout sample (B04).  The observed $i$-dropouts are both smaller
and of higher surface brightness than the cloned $U$-dropout sample (a
$2.9_{-0.8} ^{+1.3} \times$ evolution in size, see Figure 1).}
]
\end{figure}

\begin{figure}[h]
\epsscale{0.95}
\plotone{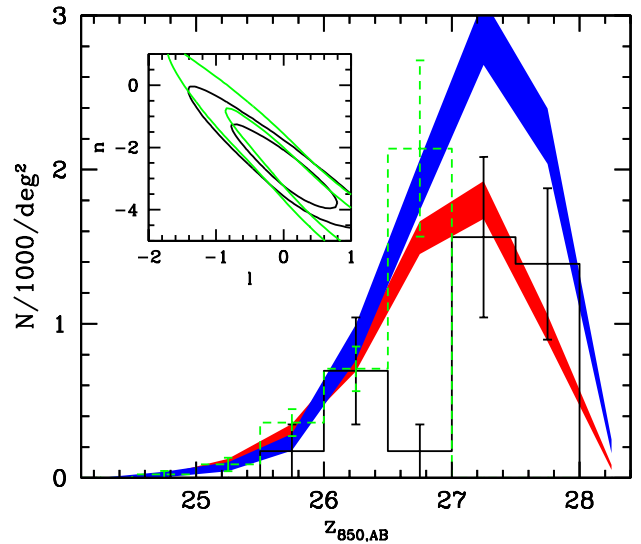}
\caption{The observed number of $i$-dropouts per half magnitude
interval observed in the UDF-Ps (solid black histogram shows the
counts from the data degraded to match the projected PSF of the lower
redshift samples [the undegraded data have an additional 8 objects
beyond 27.5, see Table 1]) versus that predicted by a luminosity
(blue-shaded region) and density (red-shaded region) scaling of the
lower redshift $B$-dropout population.  Each was scaled in size by
$(1+z)^{-1.5}$ to match the evolution found in Figure 1 (the size
scaling increased the $B$-dropout count predictions by 30-100\%).  The
green histogram shows the number of $i$-dropouts per half magnitude
interval observed in the GOODS fields using similar selection criteria
(B04) corrected up to the completeness level of the UDF-Ps.  The
figure inset shows the 68\% and 95\% confidence intervals for
luminosity evolution (parametrized as $(1+z)^l$) and density evolution
(parametrized as $(1+z)^n$) taking as inputs the GOODS data alone
(\textit{green contours}) and the GOODS plus UDF parallel data
(\textit{black contours}).  While the observations (bright
$i$-dropouts from GOODS and faint $i$-dropouts from the UDF-Ps) prefer
a model where $l=0.1$ and $n=-2.8$, pure luminosity evolution or pure
density evolution cannot be excluded (at 95\% confidence).}
\end{figure}

\begin{figure}[h]
\epsscale{1.05}
\plotone{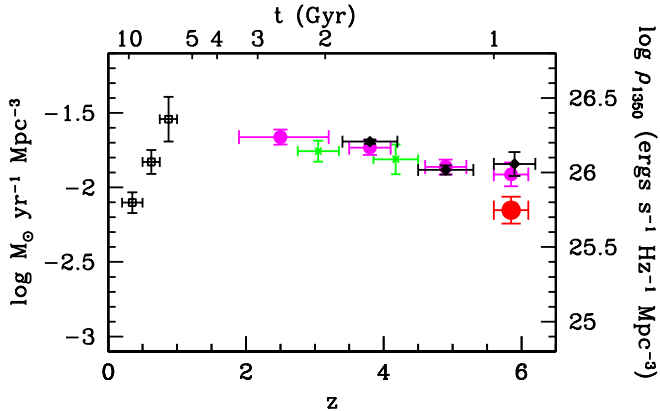}
\caption{A history of the star formation rate density assuming no
extinction correction, integrated down to $0.2L_{*}$.  The red point
at $z\sim5.9$ represents our estimate based upon the $i$-dropouts
found in both GOODS and the UDF-Ps (\S3(c)).  A Salpeter IMF is used
to convert the luminosity density into a star formation rate (e.g.,
Madau et al.\ 1998).  Comparison is made with the previous high
redshift determinations of Lilly et al.\ (1996) (open squares),
Steidel et al.\ (1999) (green crosses), Giavalisco et al.\ (2004)
(solid black diamonds), and B04 (solid magenta circles).  The top
horizontal axis provides the corresponding age of the universe.  The
uncertainties for all determinations are formal errors.  The inclusion
of size evolution makes the present estimates lower than previous
$z\sim6$ estimates.}
\end{figure}

\textit{(d) Luminosity Function.} It is also useful to construct a
simple estimate of the LF to examine the shape and look for evolution
in the faint end slope.  For our procedure, we follow Bouwens et al.\
(2003b) and treat the determination of the luminosity function
$\phi(M)$ as a deconvolution problem on the integral $\int
\phi(M(m,z)) p(m,z) \frac{dV}{dz} dz = N(m)$ where $p(m,z)$ is the
selection efficiency (including various factors ranging from galaxy
color to photometric scatter to detectability), $N(m)$ is the number
counts, and the absolute magnitude $M$ here is given as a function of
the apparent magnitude $m$ and the redshift $z$ (using an average
color dropout to perform the $k$-correction).  Adopting $N(m)$ from
the primary data sets and taking $p(m,z)$ from the $B$-dropout cloning
simulations (scaled in size as $(1+z)^{-1.5}$) (uncertain on the 20\%
level due to uncertainties in the size distribution), we are able to
compute the LF and provide a simple fit to the Schechter function
(Figure 5).  Note that the relative normalization of the bright end of
the LF ($M_{1350,AB}<-19.7$), determined primarily from GOODS, and the
faint end of the LF, derived from the UDF-Ps, could vary by 36\% RMS
(the expected cosmic variance for a CDM-type power spectrum normalized
to the high redshift observations, see B04), affecting the shape of
the LF.

\begin{figure}[h]
\epsscale{0.95}
\plotone{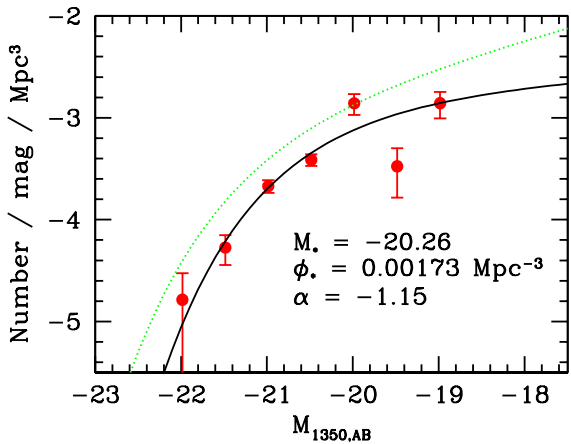}
\caption{The rest-frame continuum $UV$ ($\sim1350\AA$) LF at $z\sim6$
determined from the GOODS fields (at $M_{1350,AB} < -19.7$) and the
UDF-Ps (red circles with $1\sigma$ error bars), with the best-fit LF
to a Schechter function also shown ($M_{1350,AB,*}=-20.26$,
$\phi_{*}=0.00173$ Mpc$^{-3}$, and $\alpha = -1.15$).  The Steidel et
al.\ (1999) $z\sim3$ LF (\textit{dotted green line}) is presented for
comparison.  A fit with $\alpha=-1.6$ (as observed at $z\sim3$,
Steidel et al.\ 1999) is also consistent (at 68\% confidence) with the
data, and yields values for $M_{1350,AB,*}$ (=$-20.77$) and $\phi_{*}$
(=0.00098 Mpc$^{-3}$) more suggestive of the interpretation given in
\S3(c) (an evolution in density).}
\end{figure}

\section{Discussion}

The additional $0.9^m$ we have in $z_{850}$ depth (and $1.1^m$ in
$i_{775}$) over the GOODS field allows us to probe further down the LF
and obtain a wider distribution of sizes at bright magnitudes than was
possible from previous data.  This depth largely compensates for the
$(1+z)^4$ cosmic surface brightness dimming, allowing us to set a more
direct constraint on object sizes at $z\sim6$ (cf. B04).  The results
here are consistent with the $(1+z)^{-1.5}$ scaling of size found at
lower redshift ($1<z<5$) for objects of fixed luminosity (Ferguson et
al.\ 2004), extending the trend to $z\sim6$.  This evolution in size
has important implications for the luminosity density inferred at this
epoch, suggesting that the star formation rate density has increased
significantly from $z\sim6$.

In this work, we present the first robust luminosity function at
$z\sim6$ and find no radical difference relative to lower redshift.
While the UDF will add depth and further constraints, the present
fields provide a crucial assessment of the cosmic variance.

\acknowledgements

We would like to thank Narciso Ben{\'i}tez, Emmanuel Bertin, Fred
Courbin, Michele de la Pena, Harry Ferguson, Andy Fruchter, Dave
Golimowski, Dan Magee, and Jon McCann for their assistance and the
referee for valuable comments.  We acknowledge the support of NASA
grant NAG5-7697 and NASA grant HST-GO09803.05-A.

{}

\begin{deluxetable}{lrrrrrrrrrr}
\tablewidth{480pt}
\tabletypesize{\footnotesize}
\tablecaption{$z\sim6$ objects from the UDF-Ps.\tablenotemark{a}\label{tbl-1}}
\tablehead{\colhead{Object ID} & \colhead{Right Ascension} &
\colhead{Declination} & \colhead{$z_{850}$} &
\colhead{$i - z$} & \colhead{S/G} &
\colhead{$r_{hl}$(\arcs)} & Sample}
\startdata
UDFP1-4109-2788 & 03:32:42.692 & -27:56:55.35 &
25.9$\pm$0.1 & 1.8 & 0.00 & 0.34 & 3\\
UDFP1-3629-862 & 03:32:49.638 & -27:56:27.47 & 26.2$\pm$0.2 & 2.0 &
0.02 & 0.26 & 3\\
UDFP1-3851-2438 & 03:32:43.958 & -27:56:43.86 & 26.3$\pm$0.1 & 1.6 &
0.18 & 0.11 & 1,2,3\\
UDFP1-4650-3354 & 03:32:40.706 & -27:57:24.28 & 26.4$\pm$0.1 & 1.7 &
0.02 & 0.18 & 1,2,3\\
UDFP1-1546-1408 & 03:32:47.047 & -27:54:48.09 & 26.9$\pm$0.1 & 1.4 & 
0.00 & 0.19 & 2\\
UDFP1-2954-1152 & 03:32:48.367 & -27:55:54.81 & 26.9$\pm$0.2 & 1.5 &
0.37 & 0.11 & 2\\
UDFP1-4040-2514 & 03:32:43.709 & -27:56:50.93 & 27.0$\pm$0.2 & 1.9 &
0.01 & 0.19 & 1,2,3\\
UDFP1-2309-1628 & 03:32:46.409 & -27:55:24.31 & 27.1$\pm$0.1 & $>$2.5
& 0.02 & 0.11 & 1,2,3\\
UDFP1-4784-3382 & 03:32:40.634 & -27:57:31.07 & 27.2$\pm$0.2 & 1.6 &
0.02 & 0.14 & 1,2,3\\
UDFP1-3350-3425 & 03:32:40.111 & -27:56:22.32 & 27.4$\pm$0.2 & 1.5 &
0.01 & 0.12 & 1,2,3\\
UDFP1-4413-3612 & 03:32:39.671 & -27:57:13.38 & 27.4$\pm$0.2 & $>$2.1
& 0.02 & 0.11 & 1,2,3\\
UDFP1-3144-2410 & 03:32:43.876 & -27:56:08.48 & 27.4$\pm$0.2 & $>$1.8
& 0.01 & 0.19 & 1\\
UDFP1-1188-2375 & 03:32:43.512 & -27:54:33.43 & 27.5$\pm$0.2 & $>$2.2
& 0.01 & 0.10 & 1,2,3\\
UDFP1-3162-3232 & 03:32:40.791 & -27:56:12.25 & 27.5$\pm$0.2 &
$>$1.8 & 0.00 & 0.15 & 1\\
UDFP1-3407-1028 & 03:32:48.952 & -27:56:16.97 & 27.5$\pm$0.2 & 1.7 &
0.02 & 0.10 & 1,2,3\\
UDFP1-886-1917 & 03:32:44.962 & -27:54:16.93 & 27.6$\pm$0.2 & $>$1.9 &
0.01 & 0.12 & 1,3\\
UDFP1-2347-2492 & 03:32:43.360 & -27:55:29.04 & 27.6$\pm$0.2 & 1.5 &
0.02 & 0.11 & 1,2\\
UDFP1-3515-1622 & 03:32:46.745 & -27:56:24.45 & 27.7$\pm$0.2 & $>$1.8
& 0.00 & 0.14 & 1\\
UDFP1-3958-2376 & 03:32:44.206 & -27:56:46.38 & 27.7$\pm$0.1 & 2.3 &
0.07 & 0.07 & 1,2,3\\
UDFP1-1669-3099 & 03:32:40.914 & -27:54:59.91 & 27.8$\pm$0.2 & 1.4 &
0.01 & 0.11 & 1,3\\
UDFP1-2766-1640 & 03:32:46.480 & -27:55:47.12 & 27.8$\pm$0.2 & 1.7 &
0.01 & 0.11 & 1,3\\
UDFP1-4522-2133 & 03:32:45.269 & -27:57:13.66 & 27.9$\pm$0.2 & $>$1.8
& 0.24 & 0.09 & 1,3\\
UDFP1-1075-2025 & 03:32:44.801 & -27:54:26.57 & 28.0$\pm$0.2 & 1.5 &
0.01 & 0.10 & 1\\
UDFP1-2284-2177 & 03:32:44.529 & -27:55:24.81 & 28.1$\pm$0.2 & $>$1.6
& 0.01 & 0.11 & 1\\
UDFP2-1391-2218 & 03:32:06.504 & -27:48:46.86 & 26.3$\pm$0.1 & 1.6 &
0.02 & 0.21 & 1,2,3\\
UDFP2-3139-2714 & 03:32:00.000 & -27:48:27.90 & 27.0$\pm$0.2 & 1.6 &
0.09 & 0.19 & 3\\
UDFP2-1740-2767 & 03:32:05.051 & -27:48:20.66 & 27.2$\pm$0.2 & 1.6 & 
0.00 & 0.19 & 1,3\\
UDFP2-1371-4301 & 03:32:06.056 & -27:47:05.48 & 27.2$\pm$0.2 & 1.6 &
0.02 & 0.11 & 1,2,3\\
UDFP2-3571-2314 & 03:31:58.476 & -27:48:49.34 & 27.5$\pm$0.2 & 1.4 &
0.19 & 0.13 & 2\\
UDFP2-3846-4219 & 03:31:56.968 & -27:47:17.85 & 27.6$\pm$0.2 & $>$1.9
& 0.01 & 0.13 & 1\\
UDFP2-2451-4421 & 03:32:02.159 & -27:47:03.04 & 27.6$\pm$0.2 & 1.5 &
0.01 & 0.12 & 3\\
UDFP2-2600-3957 & 03:32:01.721 & -27:47:26.64 & 27.7$\pm$0.2 & $>$1.6
& 0.00 & 0.17 & 1\\
UDFP2-3047-2428 & 03:32:00.420 & -27:48:41.89 & 27.7$\pm$0.2 & $>$1.6
& 0.00 & 0.12 & 3\\
UDFP2-1637-2294 & 03:32:05.560 & -27:48:43.92 & 27.9$\pm$0.2 & 1.5 &
0.03 & 0.09 & 1\\
UDFP2-2704-2415 & 03:32:01.712 & -27:48:41.35 & 28.0$\pm$0.2 & 1.8 &
0.01 & 0.09 & 1,3\\
UDFP2-1185-3335 & 03:32:06.990 & -27:47:50.46 & 28.0$\pm$0.2 & 1.7 &
0.01 & 0.09 & 1\\
UDFP2-2634-3361 & 03:32:01.733 & -27:47:53.91 & 28.0$\pm$0.2 & $>$1.6
& 0.01 & 0.12 & 1\\
UDFP2-1897-3880 & 03:32:04.175 & -27:47:25.68 & 28.1$\pm$0.2 & 1.4 &
0.01 & 0.09 & 1
\enddata

\tablenotetext{a}{Selection criteria: $(i_{775}-z_{850})_{AB}>1.4$
plus null detection $(<2\sigma)$ in $V_{606}$.  All limits are
2$\sigma$.  ``S/G'' denotes the SExtractor stellarity parameter, for
which 0 = extended object and 1 = point source.  Several different
samples were used in the analysis.  Sample \#1 was generated by
applying our selection criteria to the undegraded images.  Sample \#2
and \#3 were generated by applying these selection criteria to the
images degraded to match the PSF of the $U$ and $B$-dropout samples,
respectively, projected to $z\sim6$ (0.14\arcs$\,$ FWHM and
0.10\arcs$\,$ FWHM).  Images from the latter samples were degraded to
a uniform S/N equivalent to 5 orbit $V_{606}$, 10 orbit $i_{775}$, and
15 orbit $z_{850}$ observations.  The union of all three samples (38
objects) is larger than the undegraded sample (sample \#1) due to
object blending and photometric scatter.  UDFP1 and UDFP2 refer to the
first and second UDF parallels, respectively.}
\end{deluxetable}


\begin{thebibliography}{}
\bibitem[Bennett et al.(2003)]{2003ApJ...583....1B} Bennett, C.~L.~et
al.\ 2003, \apj, 583, 1.
\bibitem[Bertin and Arnouts (1996)]{1996A&AS..117..393B} Bertin, E.\ and 
Arnouts, S.\ 1996, \aaps, 117, 393.
\bibitem[Blakeslee et al.(2003)]{2003adass..12..257B} Blakeslee,
J.~P., Anderson, K.~R., Meurer, G.~R., Ben{\'{\i}}tez, N., \& Magee,
D.\ 2003a, ASP Conf.~Ser.~295: Astronomical Data Analysis Software and
Systems XII, 12, 257.
\bibitem[Bouwens, Broadhurst and Silk (1998)]{1998ApJ...506..557B} Bouwens, 
R., Broadhurst, T.\ and Silk, J.\ 1998a, \apj, 506, 557.
\bibitem[Bouwens, Broadhurst and Silk (1998)]{1998ApJ...506..579B}
Bouwens, R., Broadhurst, T.\ and Silk, J.\ 1998b, \apj, 506, 579.
\bibitem[Bouwens, Broadhurst, \&
Illingworth(2003)]{2003ApJ...593..640B} Bouwens, R., Broadhurst, T.,
\& Illingworth, G.\ 2003a, \apj, 593, 640.
\bibitem[Bouwens et al.(2003)]{2003ApJ...595..589B} Bouwens, R.~J.~et al.\ 
2003b, \apj, 595, 589.
\bibitem[Bouwens et al.\ 2004]{bbi2004} Bouwens, R.J, Broadhurst,
T.J., Illingworth, G.D., Meurer, G.R., Blakeslee, J.P., Franx, M., \&
Ford, H.C.  2004, ApJ, submitted (B04).
\bibitem[Dickinson et al.(2004)]{2004ApJ...600L..99D} Dickinson, M.~et al.\ 
2004, \apjl, 600, L99.
\bibitem[Ferguson et al.(2004)]{2004ApJ...600L.107F} Ferguson, H.~C.~et 
al.\ 2004, \apjl, 600, L107 
\bibitem[Giavalisco et al.(2004)]{2004ApJ...600L.103G} Giavalisco, M.~et 
al.\ 2004, \apjl, 600, L103 
\bibitem[Lilly et al.\ 1996]{lil96} Lilly, S.J., Le Fevre, O., Hammer,
F., \& Crampton, D.  1996, \apj, 460, L1.
\bibitem[Madau et al.\ 1998]{mad98} Madau, P., Pozzetti, L. \&
Dickinson, M. 1998, \apj, 498, 106.
\bibitem[Schlegel, Finkbeiner, \& Davis(1998)]{1998ApJ...500..525S} 
Schlegel, D.~J., Finkbeiner, D.~P., \& Davis, M.\ 1998, \apj, 500, 525.
\bibitem[Sirianni et al.(2004)]{2004ApJ...583....1B} Sirianni,
M., et al.\ 2004, in preparation.
\bibitem[Somerville et al.(2004)]{2004ApJ...600L.171S} Somerville, R.~S., 
Lee, K., Ferguson, H.~C., Gardner, J.~P., Moustakas, L.~A., \& Giavalisco, 
M.\ 2004, \apjl, 600, L171.
\bibitem[Stanway, Bunker, \& McMahon(2003)]{2003MNRAS.342..439S}
Stanway, E.~R., Bunker, A.~J., \& McMahon, R.~G.\ 2003, \mnras, 342,
439.
\bibitem[Stanway, Bunker, \& McMahon(2003)]{2003MNRAS.342..439S}
Stanway, E.~R., Bunker, A.~J., \& McMahon, R.~G., Ellis, R., Treu, T.,
\& McCarthy, P.  2004, \apj, in press.
\bibitem[Steidel et al.\ (1999)]{1999ApJ...519....1S} Steidel, C.\ C., 
Adelberger, K.\ L., Giavalisco, M., Dickinson, M.\ and Pettini, M.\ 1999, 
\apj, 519, 1.
\bibitem[Thompson(2004)]{2003ApJ...596..748T} Thompson, R.~I., et al.\
2004, in preparation.
\bibitem[Yan, Windhorst, \& Cohen(2003)]{2003ApJ...585L..93Y} Yan, H., 
Windhorst, R.~A., \& Cohen, S.~H.\ 2003, \apjl, 585, L93.

\end{thebibliography}
\end{document}